\documentclass[11pt,reqno]{article}
\usepackage{amsmath}
\usepackage{amsfonts}
\usepackage{amssymb}
\usepackage{hyperref}
\usepackage{latexsym}
\usepackage[dvips]{graphicx}
\usepackage{epsf}

\allowdisplaybreaks \numberwithin{equation}{section}

\newcommand{\be}{\begin{equation}}
\newcommand{\ee}{\end{equation}}
\newcommand{\bea}{\begin{eqnarray}}
\newcommand{\eea}{\end{eqnarray}}

\newcommand{\f}{\frac}

\let\a=\alpha \let\b=\beta  \let\g=\gamma  \let\d=\delta
\let\z=\zeta     \let\th=\theta   \let\l=\lambda
\let\m=\mu    \let\n=\nu          \let\r=\rho 
\let\s=\sigma \let\t=\tau     
    
\let\G=\Gamma \let\D=\Delta   \let\L=\Lambda 
         
\let\Om=\Omega  \let\eps=\epsilon

\let\==\equiv

\newcommand{\p}{\partial}
\newcommand{\na}{\nabla}
\newcommand{\Tr}{{\rm Tr}}

\newcommand{\hb}{\bar{h}}

\newcommand{\Gb}{\bar{\Gamma}}
\newcommand{\cb}{\bar{c}}
\newcommand{\Cb}{\bar{C}}

\newcommand{\sbar}{\bar{\s}}

\newcommand{\tG}{\tilde{G}}
\newcommand{\Lt}{\tilde{\L}}
\newcommand{\Rt}{\tilde{R}}

\newcommand{\cC}{\mathcal{C}}

\newcommand{\cH}{\mathcal{H}}
\newcommand{\cL}{\mathcal{L}}
\newcommand{\cM}{\mathcal{M}}

\newcommand{\cR}{\mathcal{R}}
\newcommand{\cS}{\mathcal{S}}

\newcommand{\ug}{\underline{g}}
\newcommand{\uR}{\underline{R}}

\DeclareMathOperator{\im}{\mathrm{i}}

\begin{document}

\thispagestyle{empty}
\begin{flushright} \small
AEI-2011-043
\end{flushright}
\bigskip

\begin{center}
 {\LARGE\bfseries   Asymptotic safety goes on shell   }
\\[10mm]
Dario Benedetti
\\[3mm]
{\small\slshape
Max Planck Institute for Gravitational Physics (Albert Einstein Institute), \\
Am M\"{u}hlenberg 1, D-14476 Golm, Germany \\
{\upshape\ttfamily dario.benedetti@aei.mpg.de}
} 

\end{center}
\vspace{5mm}

\hrule\bigskip

\centerline{\bfseries Abstract} \medskip
\noindent
It is well known in quantum field theory that the off-shell effective action depends on the gauge choice and field parametrization used in calculating it.
Nevertheless, the typical scheme in which the scenario of asymptotically safe gravity is investigated is an off-shell version of the functional renormalization group equation.
Working with the Einstein-Hilbert truncation as a test bed, we develop a new scheme for the analysis of asymptotically safe gravity in which the on-shell part of the effective action is singled out and we show that the beta function for the essential coupling has no explicit gauge-dependence.
In order to reach our goal, we introduce several technical novelties, including a different decomposition of the metric fluctuations, a new implementation of the ghost sector, and a new cut-off scheme. 
We find a non-trivial fixed point, with a value of the cosmological constant which is independent of the gauge-fixing parameters.
\bigskip
\hrule\bigskip

\section{Introduction}

The renormalization group is a powerful framework for the understanding of fundamental issues in quantum field theory and its relation to critical phenomena. 
From the point of view of high energy physics, its most important success is the discovery of asymptotic freedom in gauge theories.
Building on that, Weinberg proposed an asymptotically safe scenario for quantum gravity \cite{Weinberg:1976xy,Weinberg:1980gg}, generalizing asymptotic freedom to the case in which the ultraviolet fixed point is not describing a free theory, with simple Gaussian path integral, but an interacting one. 
Lacking a small parameter to control a perturbative expansion, an assessment of whether such a scenario is realized in any theory of geometrodynamis is a much more complicated task than in the asymptotic freedom case. 
Following up on a seminal work by Reuter \cite{Reuter:1996cp}, an important amount of evidence has been collected in its favour, using continuous renormalization group methods.
The general framework of such methods is centered around an exact functional renormalization group equation (FRGE) for the scale-dependent quantum effective action, also called effective average action (EAA). 
The main approximation scheme in the quest for non-trivial fixed points is that of truncating the theory space, by which we mean specifying an ansatz for the EAA and discarding any term in the FRGE that would lead the flow out of the ansatz. The trustfulness of the approximation is tested by looking at convergence patterns in subsequent extensions of the truncation.
Many successful extensions of the early works \cite{Dou:1997fg,Reuter:2001ag,Lauscher:2001ya,Litim:2003vp} have been carried out, extending the original Einstein-Hilbert truncation to richer truncations, including polynomials in $R$ \cite{Granda:1998wn,Lauscher:2002mb,Codello:2007bd,Codello:2008vh,Machado:2007ea}, non-$f(R)$ truncations \cite{BMS1,BMS3,BGMS}, matter coupling \cite{Percacci:2003jz,BMS2} and running ghost sector \cite{Eichhorn:2009ah,Groh:2010ta,Eichhorn:2010tb}.

It is a natural question to ask, how much physical information is contained in a given truncation, and which of the couplings will enter in scattering amplitudes and physical observables. Indeed, much of our field theoretic description of physics is superfluous, as for example fields are not observables and we are free to reparametrize them without affecting the S-matrix (see \cite{Ball:1993zy} and references therein). Furthermore in gravitational and gauge theories we are used to deal with more variables than physical degrees of freedom, compensating the mismatch with a symmetry principle. When quantizing gauge theories we have to specify a prescription to dispose of the unphysical degrees of freedom, $i.e.$ we have to chose a particular gauge fixing, and we have to be careful in order to avoid a dependence of our predictions on such arbitrary choice.
A proper Faddeev-Popov procedure (or BRST symmetry) ensures exactly such independence for the physical observables.
However this is not the case for the effective action, which is not an observable per se, and in particular its renormalization group running depends on the gauge-fixing choice.
As we would like to discern the physical from the unphysical and redundant content of the effective action, without loosing the power of having a single generating functional rather than a collection of scattering amplitudes, we are led to the important question of discerning physical and unphysical content of the effective action.
Luckily there are some formal arguments that come to help us: it can be shown that any infinitesimal change of gauge-fixing function will lead to a change in the effective action which is proportional to its equations of motion \cite{Buchbinder:1992rb}. Furthermore any term which vanishes on shell does not affect the S-matrix as it is only possible to measure onÐshell S-matrix elements \cite{Ball:1993zy,Hart:1984jy,Jevicki:1987ax}. And this is also the reason why field redefinitions and gauge changes do not affect the S-matrix.

It is thanks to such arguments that one can conclude that pure gravity is renormalizable at one loop \cite{'tHooft:1974bx}, actually even finite in certain gauges \cite{Kallosh:1978wt}, and that only one out of the many possible counterterms of order $R^3$ is spoiling renormalizability at two loops \cite{Goroff:1985sz}.
One might wonder what is the gauge- and parametrization-independent content of the results on asymptotic safety, and whether we could reduce and simplify the task of testing asymptotic safety further by looking only at those components of the flow which carry true physical information.

In fact most of the arguments about field redefinitions and redundancy apply also to the flow of the EAA.
Under an infinitesimal change of the running scale $k\to k+\d k$, the effective average action changes as $\G_k \to \G_k + \d k\, \p_k \G_k $. If the variation contains a part proportional to the equations of motion, $ \p_k \G_k \sim X \cdot \d \G_k/\d\Phi + ...$, then we can compensate it with a field redefinition\footnote{Note that this is true both for the full EAA and for any truncation of it.} 
$\Phi\to \Phi + \d k X$. 
In this sense certain components of the flow are said to be redundant or inessential.
As a general effective average action will be parametrized by a set of running couplings $\{g_i(k)\}$,
in order to disentangle redundant from non-redundant components of the flow it is useful to introduce the notion of essential and inessential couplings \cite{Weinberg:1980gg}.
An inessential coupling is one whose flow can be compensated by a field redefinition. Equivalently, we can say that a coupling $g_j$ is inessential if and only if $\p \G_k /\p g_j$ vanishes on shell. All the other couplings are said essential.\footnote{Note that often this definition requires also some smart reparametrization of the couplings, as it can be that for example two given couplings appear essential at first glance, whereas only their ratio is actually so. The Einstein-Hilbert action in Sec.~\ref{Sec:general} will provide an explicit example of that.}
In the case of gravity there is actually a subtlety concerning the fact that a field redefinition, being a redefinition of the metric tensor, also redefines the running scale $k$. It was argued in \cite{Percacci:2004sb,Percacci:2004yy} that beacause of this the running of the (technically inessential) Newton's constant cannot be frozen and we should require the existence of a fixed point also for it.
We will see explicitly how this affects the flow equation.

The present work is motivated by the desire to isolate the on-shell component of the RG flow of gravity, which we would expect to be gauge- and parametrization-independent.
For this reason we will construct an on-shell expansion of the flow equation. We will work with the good old Einstein-Hilbert truncation, which has been used also in previous studies of the gauge dependence of FRGE for gravity \cite{Dou:1997fg,Falkenberg:1996bq,Souma:2000vs}, but in order to achieve our goal we will introduce a number of novelties with respect to standard treatments. In particular we will adopt a different decomposition of the metric fluctuations, a modified ghost action,
and a new type of cutoff scheme. Finally, we will evaluate the functional traces via a direct  spectral sum rather than by heat kernel expansion, so that we will be able to expand the FRGE around a solution of the equations of motion rather than around $R=0$.
At the end we will be able to check explicitly the full gauge-independence of the on-shell part of the FRGE.

In Sec.~\ref{Sec:general} we set up the basis for our construction, presenting the general philosophy with the aid of the Einstein-Hilbert truncation.
In Sec.~\ref{Sec:gauge-dec},~\ref{Sec:ghosts},~\ref{Sec:cutoff-frge}, and ~\ref{Sec:sums} we introduce step by step all our novelties.
In Sec.~\ref{Sec:FPs} we analyze the fixed-point structure obtained in our scheme, and in Sec.~\ref{Sec:concl} we conclude with a discussion of results and prospects.

\section{The general setup}
\label{Sec:general}

The Functional Renormalization Group Equation (FRGE)  \cite{Wetterich:1992yh,Reuter:1993kw} takes the generic form
\be\label{FRGE}
\p_t \Gamma_k[\underline{\Phi}, \Phi] = \f12 {\rm STr} \left[ \left( \frac{\delta^{2} \Gamma_k}{\delta \underline{\Phi}^A \delta \underline{\Phi}^B } + \cR_k \right)^{-1} \, \p_t \cR_k  \right]\\,
\ee
where $\underline{\Phi}$ denotes the collection of all the fields, $\Phi$ their background value and ${\rm STr}$ a functional supertrace. The running scale is $t=\ln k$, and $\cR_k$ is the cutoff function. For further details we refer to the many general reviews \cite{Morris:1998da,Bagnuls:2000ae,Berges:2000ew,Gies:2006wv,Pawlowski:2005xe,Rosten:2010vm} and to the gravity-oriented ones \cite{Niedermaier:2006wt,Reuter:2007rv,Litim:2008tt,Percacci:2007sz}.

In the case of pure gravity, the fields comprise the metric, the ghosts and occasionally some auxiliary fields implementing the functional Jacobians originated by field redefinitions.
We use the background field method to obtain a gauge-invariant AEE, and we define the decomposition of the metric by
\be \label{backgr}
\underline{g}_{\m\n} = g_{\m\n}+h_{\m\n} \, ,
\ee
with $g_{\m\n}$ denoting the background and $h_{\m\n}$ the fluctuations. For the background metric we will take that of a $d$-dimensional sphere, which is sufficient for the study of our truncation.

Following  \cite{Reuter:1996cp}, it is useful to cast a general truncation of the effective average action into the form
\be
\G_k[\underline{\Phi},\Phi] = \Gb_k[\ug] + \widehat{\G}_k[h, g] + 
\G_{\rm gf}[h, g] + \G_{\rm gh}[h, g, {\rm ghosts}] + S_{\rm aux}[g, {\rm aux.fields}] \, .
\ee
 In this decomposition $\bar{\Gamma}_k[\ug]$ depends only on the total metric. $\G_{\rm gf}$ and $\G_{\rm gh}$ denote the gauge-fixing and ghost-terms respectively, for which we will take the classical functionals but eventually allowing a running of the gravitational couplings, while $S_{\rm aux}$ is a coupling-independent action encoding the Jacobians. $\widehat{\Gamma}_k[h, g]$ encodes the explicit dependence on the background metric, it vanishes for $h=0$, 
 and it captures the quantum corrections to the gauge-fixing term. The role of the latter has been investigated via bimetric truncations in \cite{Manrique:2009uh,Manrique:2010am}; in the present work we will make use of the approximation $\widehat{\Gamma}_k=0$.

We will use our favorite test truncation, the Einstein-Hilbert truncation,
\be
\Gb_k[\ug] = Z_k \int d^d x \sqrt{\ug} \left( 2 \L_k-\uR\right) = \f{1}{16\pi} \int d^d x \sqrt{\ug} \left( 2 \f{\t_k}{G_k^{\f{d}{d-2}}}-\f{\uR}{G_k} \right)  \, ,
\ee
where all the $k$-dependence is implicitly encoded in the coupling constants, and we have introduced 
\be
Z_k = \f{1}{16\pi G_k} \, ,
\ee
and the essential coupling
\be \label{tau}
\t_k = G_k^{\f{2}{d-2}} \L_k \, .
\ee
In this parametrization of the couplings, the left-hand-side of \eqref{FRGE} (projected as usual on the background) reduces to
\be \label{lhs1}
\p_t \Gb_k[g]  = \p_t \t_k  \f{1}{8\pi G_k^{\f{d}{d-2}}} \int d^d x \sqrt{g}  + \f{\p_t G_k }{G_k} \f{1}{16\pi} \int d^d x \sqrt{g} \left( -\f{2d}{d-2} \f{\t_k}{G_k^{\f{d}{d-2}}}+\f{R}{G_k} \right) \, .
\ee
The interesting thing is that the piece with the anomalous dimension $\eta \equiv \f{\p_t G_k }{G_k}$ is proportional to the equations of motion
\be \label{eom}
R = \f{2 d}{d-2} \L_k \, ,
\ee
a simple restatement of the fact that $G_k$ is an inessential parameter.
We could also choose a parametrization in terms of $\t_k$ and $\L_k$, such that the role of inessential parameter be played by  the cosmological constant,
\be
\Gb_k[\ug] = \f{\t_k^{\f{2-d}{2}}}{16\pi} \int d^d x \sqrt{\ug} \left( 2 \L_k^{\f{d}{2}}-\L_k^{\f{d-2}{2}} \uR\right) \, ,
\ee
as well as several other possibilities \cite{Percacci:2004yy}, involving in particular also redefinitions of the essential parameter. Eventually one would like to ``define the coupling constants as coefficients in a power series expansion of the reaction rates themselves around some physical renormalization point'' \cite{Weinberg:1980gg}, but we will content ourselves with distinguishing one essential and one inessential coupling.
Most of our results and discussions are independent of the coupling parametrization, and the choice $(\t_k,G_k)$ fits well enough to our purposes.

As argued in the introduction, we are interested in isolating the gauge-independent information contained in the FRGE, which we know should be associated with the on-shell part of the equation. For the lhs of \eqref{FRGE} it is very easy to go on shell, and only the first term of \eqref{lhs1} survives.
For the right-hand-side things get more difficult, and it is for this reason that we are going to introduce some new methods in the following sections.
In the standard approach, the rhs is expanded around $R=0$, with the functional traces being evaluated via a heat kernel expansion; beta functions are then obtained equating the coefficients of equal powers of $R$ in the left- and right-hand-side of the equation. If we wish to go on shell, the problem in applying this standard strategy is that, for non-zero cosmological constant, all powers of $R$ will contribute to the on-shell part. That is, we would have to know the whole series generated by the heat kernel expansion and be able to resum its on-shell part. As this seems a rather formidable task, the only consistent way to use the heat kernel on shell is to also expand to the same order in $\L$, as indeed was done in \cite{Dou:1997fg}.
Our strategy for the rhs of \eqref{FRGE} will be different: rather than employing the heat kernel expansion, we will compute the traces via an explicit spectral sum over a spherical background.
In this way we will retain all powers of $R$, and we will be able to expand around the solution of \eqref{eom}, without having to resum an infinite (and not known to all orders) series from the heat kernel expansion.

Expanding to first order in $R - \f{2 d}{d-2} \L_k$ we will obtain a coupled system
\bea \label{system-a}
\p_t \t_k &=& \b_\t (\t_k,\eta, \tG_k) \, ,\\ \label{system-b}
\eta &=& \g (\t_k, \tG_k) \, ,
\eea
where $\tG_k \equiv k^{d-2} G_k$ is the dimensionless Newton constant.
Its appearance as an independent argument in $\b_\t$ and $\g$ is what spoils its full interpretation as an inessential parameter, and it is a manifestation of the double role played by the metric (in defining scales and being a dynamical field) which generically makes it impossible to remove Newton's constant from the RG equations \cite{Percacci:2004sb,Percacci:2004yy}.
If this was not the case, we would have a standard situation, with the anomalous dimension  $\eta$ fixed by an algebraic equation, selecting a discrete set among the fixed point solutions $\t^*=\t^*(\eta)$ of $\b(\t,\eta)=0$.
Because of the explicit dependence on $\tG_k$, it appears natural to switch to the formulation common in asymptotically safe gravity, in which we treat both $\Lt_k$ and $\tG_k$ as essential couplings:
\bea \label{system2-a}
\p_t \Lt_k &=& \b_{\Lt} (\Lt_k,\tG_k) \, , \\  \label{system2-b}
\p_t \tG_k &=& (d-2 +\eta(\Lt_k,\tG_k)) \tG_k \, .
\eea
Despite that, we will see that the decomposition into essential and inessential couplings is nevertheless very useful, and it will help us in isolating the gauge-independent component from the gauge-dependent one.

\section{Hessian, symmetries and gauge-invariant decomposition}
\label{Sec:gauge-dec}

The background decomposition \eqref{backgr} leads to an expansion of the action of the type
\be \label{hTaylor}
\Gb[g+h] = \Gb[g] + \Gb^{(1)}_{\m\n}\cdot h^{\m\n} + \f12 h^{\m\n}\cdot \Gb^{(2)}_{\m\n,\r\s} \cdot h^{\r\s} + ...
\ee
where $\Gb^{(n)}$ stands for the $n$-th functional derivative with respect to $\ug_{\m\n}$, evaluated at $\ug_{\m\n}=g_{\m\n}$,
and $\cdot$ stands for an integration.

From the lhs of \eqref{hTaylor},
it is obvious that the action has two type of symmetries:\footnote{The action is invariant under diffeomorphisms. Remember that for an infinitesimal transformation $x^\m\to x^\m+\eps^\m$, the metric transforms as
$\underline{g}_{\m\n}\to\underline{g}_{\m\n}+\cL_\eps\underline{g}_{\m\n}=\underline{g}_{\m\n}+\cL_\eps g_{\m\n}+\cL_\eps h_{\m\n}$, 
where $\cL_\eps\underline{g}_{\m\n} = \eps^\a \p_\a\underline{g}_{\m\n} +\underline{g}_{\a\n} \p_\m\eps^\a +\underline{g}_{\a\m} \p_\n\eps^\a$ is the Lie derivative, which is a linear operator. 
Note also that, with the aid of the background-covariant derivative, we can rewrite  $\cL_\eps g_{\m\n} = \na_\m \eps_\n +\na_\n \eps_\m$,  and
 $\cL_\eps h_{\m\n} = \eps^\a \na_\a h_{\m\n} +h_{\a\n} \na_\m\eps^\a +h_{\a\m} \na_\n\eps^\a$} 
\begin{enumerate}
\item Invariance under independent tensorial transformations of $g_{\m\n}$ and $h_{\m\n}$:
\bea \label{symm1}
g_{\m\n} &\to & g_{\m\n} + \cL_\eps g_{\m\n} \, ,\\
h_{\m\n} &\to & h_{\m\n} + \cL_\eps h_{\m\n} \, .
\eea

\item Invariance under gauge-type transformations on a fixed background:
\bea
g_{\m\n} &\to & g_{\m\n}  \, ,\\   \label{symm2}
h_{\m\n} &\to & h_{\m\n} + \cL_\eps (g_{\m\n}+h_{\m\n}) \, .
\eea
\end{enumerate}

While the first transformation is trivially implemented, order by order, also on the rhs of the expansion, one has to be more careful with the second type of transformation.
Indeed, as the transformation of $h_{\m\n}$ in \eqref{symm2} contains an $h$-independent term, different orders in the expansion will mix under the transformation.
In particular, at first order in $h_{\m\n}$, it is the combination
\be
\Gb^{(1)}_{\m\n}\cdot  \cL_\eps h^{\m\n} +  h^{\m\n}\cdot  \Gb^{(2)}_{\m\n,\r\s}\cdot  \cL_\eps g_{\r\s} 
\ee
which is zero. Only on shell, when $\Gb^{(1)}_{\m\n}\equiv \f{\d \Gb}{\d g^{\m\n}} = 0$, the second-order part of the action is invariant on its own under
\be \label{h-symm}
h_{\m\n} \,\, \to \,\, h_{\m\n} + \cL_\eps g_{\m\n} \, .
\ee
And only on shell will the Hessian contain zero modes rendering it not-invertible.
In this sense, in perturbative calculations it is only the need to go on shell that forces us to introduce a gauge-fixing term.
On the other hand the term $h^{\m\n}\cdot  \Gb^{(2)}_{\m\n,\r\s}\cdot  \cL_\eps h_{\r\s} $ is different from zero also on shell, and it only cancels in combination
with the term of the same order coming from the next term in the Taylor expansion \eqref{hTaylor}. The story goes on in a similar fashion for the higher-order terms.
Because of the 1-loop structure of the FRGE, and because we set $h=0$ in its evaluation, discarding any non-trivial running of the gauge-fixing and ghost terms,
it is only the on-shell symmetry of the Hessian that plays an important role in our analysis, and 
for this reason, in the following we will only be concerned about the on-shell symmetry \eqref{h-symm}.
Thanks to that, we will now be able to switch to gauge-invariant and pure-gauge variables.

The transverse-traceless decomposition of the metric tensor is given by
\be \label{TT-dec}
h_{\mu\nu} = h_{\mu\nu}^{T} + \na_\m \xi_\n + \na_\n \xi_\m + \na_\mu \na_\nu \s - \frac{1}{d} g_{\mu\nu} \na^2 \s + \frac{1}{d} g_{\mu\nu} h\, ,
\ee
with the component fields satisfying
\be
g^{\mu \nu} \, h_{\mu\nu}^{T} = 0 \, , \quad \na^\mu h_{\mu\nu}^{T} = 0
\, , \quad \na^\mu \xi_\mu = 0 \, , \quad h = g_{\mu \nu} h^{\mu \nu} \, .
\ee
Decomposing also the diffeomorphism parameter as $\eps_\m = \eps^T_\m +\na_\m\eps$ (with $\na^\m\eps^T_\m=0$), the symmetry \eqref{h-symm} is rewritten as
\bea
h^T_{\m\n} &\to &  h^T_{\m\n} \, , \\
 \xi_\m &\to  & \xi_\m + \eps^T_\m \, , \\
 \s  &\to & \s + 2\eps  \, ,\\
 h  &\to & h + 2 \na^2 \eps \, .
\eea
It is appropriate to introduce the gauge-invariant scalar
\be
\hb = h +\D \s \, ,
\ee
where we also defined $\D =-\na^2$.

We will also redefine the field $\s$, to bring the gauge-fixing part of the action into a simple form, and completely decouple (on shell) the gauge-invariant from the gauge-variant field components.
To this end, we recall that a general gauge-fixing term for the type of truncation we are considering here is
\be
\G_{\rm gf}[h, g] = \f{Z_k}{2\a} \int d^d x\sqrt{g} F_\m F^\m \, ,
\ee
with
\be
F_\m \equiv \na^\n h_{\m\n} - \f{1+\b}{d} \na_\m h  = -\Big(\D -\f{R}{d}\Big) \xi_\m -\na_\m \Big(\f{d-1-\b}{d}\D-\f{R}{d}\Big)\s -\f{\b}{d} \na_\m \hb \, .
\ee
The parameter $\a$ determines the width of the Gaussian fluctuations around the gauge-fixed surface $F_\m=0$, while $\b$ parametrizes different choices of $F_\m$.
We define the new field
\be \label{sbar}
\sbar = \s + \f{\b}{(d-1-\b)\D-R} \hb \, ,
\ee
in terms of which it becomes manifest the dependence of the gauge-fixing term on only two fields:
\be
\G_{\rm gf}[h, g] = \f{Z_k}{2\a} \int d^d x\sqrt{g} \left\{ \xi^\m \Big(\D-\f{R}{d}\Big)^2 \xi_\m + \sbar\Big(\f{d-1-\b}{d}\D-\f{R}{d}\Big)^2\D\sbar      \right\} \, .
\ee
We have achieved a separation between the gauge-invariant fields $\{h^T_{\m\n},\hb\}$ and the gauge-variant ones $\{\xi_\m,\sbar\}$.
Note that this is true only as long as we discard the higher order part of \eqref{symm2}, i.e. the $\cL_\eps h_{\m\n}$ term,
and this is an approximation compatible with the truncation of the Taylor expansion \eqref{hTaylor} to second order in $h_{\m\n}$.

The change of variables \eqref{TT-dec} gives rise to a non-trivial Jacobian, which will be accounted for in Sec.~\ref{Sec:ghosts} with the introduction of auxiliary fields \cite{Machado:2007ea,Codello:2008vh}. On the contrary, the field redefinition $\{h,\s\}\to\{\hb,\sbar\}$ has a trivial Jacobian and brings no other auxiliary fields in the action.

\section{Ghosts and auxiliary fields}
\label{Sec:ghosts}

In order to discuss our peculiar choice for the ghost sector we have to anticipate some aspects of the results of our full calculation.
With the usual implementation of the ghost sector, we found that the on-shell traces depend on $\eta_\a\equiv -\f{\a}{Z_k} \p_t \f{Z_k}{\a} = \eta +\f{\p_t\a}{\a}$, which carries with it a gauge-dependence.
The typical choice is to assume $\p_t\a=0$ but to keep the running of $Z_k$, $i.e.$ $\eta_\a=\eta$, which amounts to dictating a specific running of the gauge-fixing term, with $\f{Z_k}{k^2\a}$ eventually reaching a fixed point\footnote{It is expected, in analogy to Yang-Mills \cite{Ellwanger:1995qf} and from general arguments \cite{Litim:1998qi}, that $\a=0$ should be a fixed point. Nevertheless this does not fix the anomalous dimension $\eta_\a$ at $\a=0$, as no argument is known which can fix the undetermined zero over zero  $\f{\p_t\a}{\a}$ without an explicit calculation.} together with $\tG_k$. 
As we are not computing the beta function for $\a$, other choices for $\eta_\a$ would be equally valid, and we can view any given choice of $\eta_\a$ as part of the scheme dependence of the FRGE in our truncation.

Here we want to suggest that an exact cancellation between pure-gauge and ghost degrees of freedom should be realized on-shell. Note that such a cancellation was achieved (off-shell, but in a special gauge) in higher derivative theories \cite{Codello:2008vh,Machado:2007ea,BMS2}, where effectively the equivalent of $\eta_\a$ was set to zero. We want to achieve it on-shell also for general gauge and $\eta_\a$,\footnote{We will always assume $\p_t\b=0$, but the cancellation mechanism we propose works also for terms proportional to $\p_t\b$.} and for this reason we propose a different implementation in terms of ghosts of the Faddeev-Popov determinant.

First of all, one notes that in full rigor an overall factor $\sqrt{\f{Z_k}{\a}}$ should be present in the ghost action (one way to see this is that we have a factor $\sqrt{Z_k}$ from the constraint functional, and a factor $1/\sqrt{\a}$ from the so called third ghost part \cite{Buchbinder:1992rb}). Since such a factor is field- and derivative-independent, it is usually discarded in perturbative calculations, as well as in the literature on asymptotically safe gravity.
The justification for discarding it in the analysis of the FRGE is that if we discard its running, then it completely disappears from the FRGE. However, if we let it run, it will contribute to the FRGE with terms proportional to $\eta_\a/2$, and as $\eta_\a$ is usually not discarded in the gauge-fixing sector, we find its omission from the ghost contribution not fully consistent\footnote{Of course, as we are not deriving beta functions for the gauge-fixing and ghost terms, the choice to discard or not the running of such terms is part of the arbitrariness of the truncation. 
Note that in \cite{Groh:2010ta,Eichhorn:2010tb} the running of the ghost kinetic term was studied, but not that of the gauge parameter $\a$, so the extent to which the ghost anomalous dimension differs from the anomalous dimension $\eta_\a$ of the gauge-fixing term is an open question. In the present work we assume that they are related exactly as in the path integral construction and we denote the anomalous dimension for both the ghost and gauge-fixing sector with $\eta_\a$.}.
The inclusion of  a factor $\sqrt{\f{Z_k}{\a}}$ turns out not to be the end of the story: we found that its inclusion does not eliminate the dependence of the on-shell equation on $\eta_\a$.

In order to achieve our goal, remember that, calling $\cM$ the ghost (or Faddev-Popov) operator, the Faddev-Popov determinant is $|\det (\cM) |$, and it is this which we usually  implement via a path integral over the ghosts.
However, barring a multiplicative anomaly, the Faddeev-Popov determinant is also equivalent to  $\sqrt{\det (\cM^2)} $. In a path integral we can realize these two expressions respectively as
\be \label{ghosts1}
\G_{\rm gh,1} = \sqrt{\f{Z_k}{\a}} \int d^d x \sqrt{g} \left\{ \Cb^{T \m} \Big( \D -\f{R}{d}\Big) C_\m^T +2 \cb \Big( \f{d-1-\b}{d} \D -\f{R}{d}\Big) \D c    \right\} \, ,
\ee
and
\be \label{ghosts2}
\begin{split}
\G_{\rm gh,2} = \f{Z_k}{\a} \int d^d x \sqrt{g} \Big\{ & \Cb^{T\m} \Big( \D -\f{R}{d}\Big)^2 C_\m^T +4 \cb \Big( \f{d-1-\b}{d} \D -\f{R}{d}\Big)^2 \D c    \\
   & +  B^{T\, \m} \Big( \D -\f{R}{d}\Big)^2 B_\m^T +4 b \Big( \f{d-1-\b}{d} \D -\f{R}{d}\Big)^2 \D b \Big\}\, ,
\end{split}
\ee
where $C_\m^T$ and $c$ are complex Grassmann fields, while $B^T_\m$ and $b$ are real bosonic fields, and the vectors are all transverse.
In both cases, the functional integration over the ghosts\footnote{In analogy with the ``third ghost'' of higher derivative gravity \cite{Buchbinder:1992rb}, we keep the name ghost also for $B^T_\m$ and $b$, even if they are normal bosonic fields.} exactly cancels with the integral over $\xi$ and $\sbar$ of $e^{-S_{\rm gf}}$.
On the other hand, implementing a cutoff via the rule \eqref{cutrule}, one finds that the relative functional traces appearing on the rhs of \eqref{FRGE} do not cancel in the first case (unless $\eta_\a=0$), while they exactly cancel in the second.
In order to achieve full gauge-independence of the on-shell equation, we will adopt in the following the new ghost action \eqref{ghosts2}.\\

The final ingredient of our truncation is the action for the auxiliary fields, introduced to take into account the Jacobian arising in the TT decomposition \eqref{TT-dec}.
The Jacobian for the gravitational sector leads to the standard auxiliary action
\be \label{aux-gr}
\begin{split}
S_{\rm aux-gr} =  \int d^d x \sqrt{g} \Big\{ & 2 \bar{\chi}^{T\, \m} \Big( \D -\f{R}{d}\Big) \chi_\m^T + \bar{\chi} \Big( \f{d-1}{d} \D -\f{R}{d}\Big) \D \chi    \\
   & + 2 \z^{T\m} \Big( \D -\f{R}{d}\Big) \z_\m^T + \z \Big( \f{d-1}{d} \D -\f{R}{d}\Big) \D \z  \Big\}\, ,
\end{split}
\ee
where the $\chi_\m^T$ and $\chi$ are complex Grassmann fields, while $\z_\m^T$ and $\z$ are real fields.
The Jacobian for the transverse decomposition of the ghost action, in its new version \eqref{ghosts2}, is given by
\be
S_{\rm aux-gh} = \int d^d x \sqrt{g} \, \phi \D \phi \, ,
\ee
which differs from the one we would obtain from a standard ghost sector  \eqref{ghosts1} \cite{Codello:2008vh,Machado:2007ea} only in the fact that the field $\phi$ is real rather than complex.

\section{On-shell cutoff scheme}
\label{Sec:cutoff-frge}

We use the standard cutoff scheme (of Type I, in the nomenclature of \cite{Codello:2008vh}), which amounts to choosing the cutoff $\cR_k$ in such a way to implement in the Hessian the rule
\be \label{cutrule}
\D\to P_k\Big( \f{\D}{k^2} \Big) \equiv \D+k^2 r_k\Big( \f{\D}{k^2} \Big) \, ,
\ee
where $r_k(x)$ is some fixed cutoff profile function.
However, we do not implement it on all the terms of the gauge-fixed Hessian. Rather, we apply it only to those terms that survive when the background goes on-shell.
It is clear that such a choice is valid, as we are still enforcing a cutoff on the low-modes for every single field in the path integral, thanks to the presence of the gauge-fixing term.
This choice, which we will dub ``on-shell Type I'', is an important step of our construction, which helps separating gauge-invariant from gauge-variant fields in the FRGE, and facilitates its analysis.

The quadratic part of the gauge-fixed gravitational action, can be decomposed as
\be \label{hessian}
\begin{split}
\Big(  \f12 h^{\m\n}\cdot \Gb^{(2)}_{\m\n,\r\s} \cdot h^{\r\s} + \G_{\rm gf}   & \Big)_{|_{S^d}}   = \\
   = Z_k \int d^d x\sqrt{g} \Big\{ & \f14 h^{T\m\n} \Big(\D+\f{2}{d(d-1)}R \Big) h^T_{\m\n} - \f{d-2}{4d^2} \hb \Big( (d-1) \D -R\Big) \hb \\
   & +\f{1}{2\a} \xi^\m \Big( \D-\f{R}{d} \Big)^2 \xi_\m + \f{1}{2d^2\a} \sbar \Big( (d-1-\b) \D -R\Big)^2 \D\sbar \\
   & +  \Big(R - \f{2 d}{d-2} \L_k\Big) X \Big\}  \, , 
\end{split}
\ee
with
\be
\begin{split}
X = &\, \f{d-2}{4d} h^{T\m\n}  h^T_{\m\n}+ \f{d-2}{2d} \xi^\m \left( \D-\f{R}{d}\right) \xi_\m \\
  & - \f{d-2}{8d^2} \hb \Big( (d-1) \D -R\Big) \f{(d-2)\Big( (d-1)\D-R\Big)-2 \b^2\D}{\Big( (d-1-\b)\D - R \Big)^2} \hb   \\
  & + \f{d-2}{8d^2} \sbar \Big( d \D -2 R \Big) \D \sbar + \f{d-2}{4d^2} (d-2-2\b) \hb  \f{ (d-1)\D-R}{ (d-1-\b)\D - R } \D\sbar \, .
\end{split}
\ee
We see that for $\b\neq 0$ the definition \eqref{sbar} has led to a non-local off-shell Hessian, which reduces to a local one on shell.
It is only the latter that we wish to regulate in our scheme. That is, it is only on the first four terms of \eqref{hessian} that we impose the cutoff rule \eqref{cutrule}, leaving the part proportional to the equations of motion $R - \f{2 d}{d-2} \L_k$ untouched.
The ghost and auxiliary sectors do not have such a natural decomposition, and the rule applies as in the standard case.

In practice, we have
\be \label{cutrule2}
\cR_k^{(\Phi_i \Phi_j)} = \cH_{\rm on-shell}^{(\Phi_i \Phi_j)}(P_k)  - \cH_{\rm on-shell}^{(\Phi_i \Phi_j)}(\D) \, ,
\ee
for any term of the type $\f12 \Phi_i \cH^{(\Phi_i \Phi_j)}(\D) \Phi_j$ in the quadratic part of the truncated EAA, where by $\cH_{\rm on-shell}$ we mean that part of the Hessian that survives on shell.
For example, we have
\be
\cR_k^{(\xi\xi)} = \f{Z_k}{\a} \Big( P_k-\f{R}{d} \Big)^2 - \f{Z_k}{\a} \Big( \D-\f{R}{d} \Big)^2 \, ,
\ee
for the $\xi\xi$ part, and so on.

We omit the details, just pointing out that one of the simplifications of our choice of variables and cutoff scheme is to diagonalize the $(\hb,\sbar)$ matrix being traced in the FRGE. Indeed from \eqref{hessian} and \eqref{cutrule2} it is clear that no cutoff is introduced for the mixed $(\hb,\sbar)$ terms.

\section{On-shell expansion and spectral sums}
\label{Sec:sums}

As discussed in Sec.~\ref{Sec:general}, we are going to evaluate the traces by a direct spectral sum rather than in a heat kernel expansion.
By spectral sum\footnote{Spectral sums have been used before for a conformally reduced truncation \cite{Reuter:2008qx}, and for topologically massive gravity \cite{Percacci:2010yk}.} we mean that a generic trace will be evaluated as
\be \label{eigsum}
\Tr_s W(\D) = \sum_n D_{n,s} W(\l_{n,s}) \, ,
\ee
where $\{\l_{n,s}\}$ is the spectrum of eigenvalues of the Laplacian $\D$ on spin-$s$ fields, with the relative multiplicities $\{D_{n,s}\}$.
The spectrum of the Laplacian on a $d$-dimensional sphere can be found in \cite{Rubin:1983be} (see also \cite{Lauscher:2001ya,Codello:2008vh}), and we report it for convenience in Table~\ref{table}.
\begin{table}
\begin{center}
\begin{tabular}{|c|c|c|}\hline
Spin s  & Eigenvalue $\l_{n,s}$ & Multiplicity $D_{n,s}$\\\hline
0 & $\frac{n(n+d-1)}{d(d-1)}R$; $n=0,1\ldots$& $\frac{(n+d-2)!\, (2n+d-1)}{n!(d-1)!}$\\\hline
1 & $\frac{n(n+d-1)-1}{d(d-1)}R$; $n=1,2\ldots$& $\frac{(n+d-3)!\, n(n+d-1)(2n+d-1)}{(d-2)!(n+1)!}$\\\hline
2 & $\frac{n(n+d-1)-2}{d(d-1)}R$; $n=2,3\ldots$& $\frac{(n+d-3)!\, (d+1)(d-2)(n+d)(n-1)(2n+d-1)}{2(d-1)!(n+1)!}$\\\hline
\end{tabular}\end{center}
\caption{Eigenvalues of the Laplacian on the $d$-sphere and their multiplicities}
\label{table}
\end{table}

We have to be careful not to include fictitious modes in the sum \cite{Codello:2008vh,Machado:2007ea}. 
Remembering that our decomposition for the metric fluctuations is
\be \label{inv-TT-dec}
h_{\mu\nu} = h_{\mu\nu}^{T} + \na_\m \xi_\n + \na_\n \xi_\m + \na_\mu \na_\nu \sbar + \Big( \frac{1}{d} g_{\mu\nu} -\na_\m\na_\n \f{\b}{(d-1-\b)\D-R}\Big) \hb\, ,
\ee
we see that we should exclude two sets of modes that give no contribution to $h_{\m\n}$. First, we should exclude the Killing vectors, satisfying $\na_\m \xi_\n + \na_\n \xi_\m=0$. Second, we should leave out also the constant scalar modes $\sbar =$ constant.
A similar set of modes should be excluded also from the ghosts and auxiliary fields, as these are all fields introduced hand-in-hand with $\xi$ and $\sbar$. The only fields for which we retain all the modes are $h^T_{\m\n}$ and $\hb$.
Note that, differently from \cite{Codello:2008vh,Machado:2007ea}, we do not exclude the scalar modes corresponding to conformal Killing vectors $\cC_\m= \na_\m\sbar$, $i.e.$ those scalar modes satisfying $ \na_\mu \na_\nu \sbar =  \frac{1}{d} g_{\mu\nu} \na^2 \sbar$. It is indeed clear that in our decomposition \eqref{inv-TT-dec} such modes do contribute to $h_{\m\n}$. 
This can be seen also from the point of view of the ghosts: the ghost modes should be in one-to-one correspondence with the modes of the gauge parameter $(\eps_\m^T,\eps)$, and from $\cL_\eps g_{\m\n} = \na_\m \eps^T_\n + \na_\n \eps^T_\m + 2 \na_\mu \na_\nu \eps$ it is obvious that there is no reason to exclude the scalar modes $\eps$ corresponding to conformal Killing vectors.
We define the index of the lowest mode being included in our sums as $n_{0,s}$, for which we have $n_{0,2}=n_{0,1}=2$ and $n_{0,0}=1$. The contribution of the constant $\hb$ mode will be added separately.

We choose to work with Litim's optimized cutoff \cite{Litim:2001up}
\be
r_k(x) = (1-x) \th(1-x) \, ,
\ee
for which
\be
\p_t (k^2 r_k(\D/k^2) ) = 2 k^2 \th (k^2-\D) \, .
\ee
Its great technical advantage is that all the functions appearing in the FRGE have a numerator proportional to the step function, and hence the spectral sums are cut off at 
$N_s=\text{max} \{n\in\mathbb{N} : \l_{n,s}\leq k^2\}$. At the same time, for all $\l_{n,s}\leq k^2$, we have $P_k(\l_{n,s}/k^2) = k^2$.

Introducing the dimensionless quantities
\be
\Rt = R/k^2 \, , \,\,\,\,\,\, \Lt_k = \L/k^2 \, ,
\ee
we write the FRGE as
\be
\begin{split}
\p_t \Gb_k &=  \sum_{s=0,1,2}\, \left(\sum_{n=n_{0,s}}^{N_s(\Rt)} W_s(\l_{n,s}/k^2, \Rt, \Lt_k, \eta; \a,\b,\eta_\a) \right) + \hat{W}_{\text{const.mode-}\hb} \\
   &\equiv \cS(\Rt, \Lt_k, \eta; \a,\b,\eta_\a) \, .
\end{split}
\ee
where the functions $W_s(\D/k^2, \Rt, \Lt_k, \eta; \a,\b)$ are obtained by collecting the contributions to \eqref{FRGE} coming from all the fields of spin $s$.
Note that $N_s$ is a function of $\Rt$ as well as of the spin $s$.

As already stressed, the importance of evaluating the traces via a spectral sum is that it allows us to expand the result around a solution of the equations of motion, as we now proceed to do.
When expanding \eqref{FRGE} around the solution of \eqref{eom}, the lhs of the equation is of course the easiest part.
For our spherical background we have
\be
k^d \int d^d x\sqrt{g} = \Om_d \r^d \, ,
\ee
where
\be
\Om_d \equiv (4\pi)^{\f{d}{2}} \f{\G(\f{d}{2})}{\G(d)} \, , \,\,\,\,\, \r^2 = \f{d(d-1)}{\Rt} \, .
\ee
Substituting this formula for the volume in \eqref{lhs1}, and expanding the resulting expression around the solution of \eqref{eom}, we find
\be
\begin{split}
\p_t \Gb_k = & \p_t \t_k \f{\Om_d}{8\pi}  \left(\f{(d-1)(d-2)}{2\t_k}\right)^{\f{d}{2}} \\
   & +\left( \eta  -\f{\p_t\t_k}{d-1} \left(\f{(d-1)(d-2)}{2\t_k}\right) \right) \f{\Om_d }{16\pi} \left(\f{(d-1)(d-2)}{2\t_k}\right)^{\f{d}{2}} \tG_k^{\f{2}{d-2}} \left( \Rt - \f{2d}{d-2} \Lt_k \right)
\end{split}
\ee

For the rhs of the FRGE we have 
\be
\cS(\Rt, \Lt_k, \eta; \a,\b,\eta_\a) \simeq \cS^{(0)} (\Lt_k, \eta)+ \Big(\Rt - \f{2 d}{d-2} \Lt_k \Big) \cS^{(1)}(\Lt_k, \eta; \a,\b,\eta_\a)
\ee
where $\cS^{(0)}=\cS{|_{\Rt=\f{2 d}{d-2} \Lt_k}}$ and $\cS^{(1)}=\f{\p \cS}{\p \Rt} {|_{\Rt=\f{2 d}{d-2} \Lt_k}}$, and from now on we will omit their arguments.
The important result is that $\cS^{(0)}$ is completely gauge independent, that is, independent of $\a$, $\b$ and $\eta_\a$.

By comparing similar orders in the expansion, we obtain a system of two coupled equations
\be \label{beta-tau-d}
\p_t \t_k  = \f{8\pi}{\Om_d}  \left(\f{2\t_k}{(d-1)(d-2)}\right)^{\f{d}{2}} \cS^{(0)} \equiv \b_\t(\t_k, \eta, \tG_k) =  \tG_k^{\f{d}{d-2}} f( \Lt_k,\eta) \, ,
\ee
\be \label{eta-eq}
\tG_k = 2\eta \left( \f{(d-2) f( \Lt_k,\eta)}{\Lt_k}   +\f{32\pi}{\Om_d}\left(\f{2\Lt_k}{(d-1)(d-2)}\right)^{\f{d}{2}}  \cS^{(1)} \right)^{-1} \, ,
\ee
which is of the type we already anticipated in (\ref{system-a}-\ref{system-b}), after the substitution $\Lt_k=\t_k \tG_k^{-\f{2}{d-2}}$.

Specializing  to $d=4$, we have
\be \label{S0}
\begin{split}
\cS^{(0)} =& \frac{1}{72 \Lt_k ^2}\Bigg( \frac{3 \left(\eta  (2 \Lt_k +3)^2-72 \Lt_k ^2\right)}{4 \Lt_k -3} \\
&  -\frac{5 (2 \Lt_k -3) \left(\eta  \left(16 \Lt_k ^2+18 \Lt_k -9\right) +18
   \left(-\Lt_k +\sqrt{\Lt_k  (17 \Lt_k +12)}+3\right)\right)}{2 \Lt_k +3} \\
& +\frac{18 \left(\sqrt{\Lt_k ^3 (13 \Lt_k +12)}-38 \Lt_k ^2+9 \Lt_k +3
   \sqrt{\Lt_k  (13 \Lt_k +12)}+9\right)}{\Lt_k -1} \Bigg) \, ,
\end{split}
\ee
which can be seen to be explicitly gauge-independent.
On the contrary, $\cS^{(1)}$ is gauge-dependent and could only be evaluated analytically by specifying a numerical value for the gauge-fixing parameters $\a$ and $\b$, together with a choice for $\eta_\a$.
We report here the result for $\a=\b=0$ (for which the result is independent of $\eta_\a$):
\be
\begin{split}
\cS^{(1)} = \frac{1}{48 \Lt_k ^5} \Bigg(\frac{p_1(\Lt_k)}{(3-4 \Lt_k )^4}+\frac{p_2(\Lt_k)}{(2 \Lt_k +3)^2}+\frac{p_3(\Lt_k)}{(\Lt_k -1)^3 (13 \Lt_k +12)}   \Bigg) \, ,
\end{split}
\ee
where
\be
\begin{split}
p_1(\Lt_k) = & 3 \Lt_k ^2 \Big (-\eta  (2 \Lt_k +3) (4 \Lt_k -3)^3-4 \Big(-96 \Lt_k ^5+384 \Lt_k ^4-270 \Lt_k ^3-81 \Lt_k ^2 \\
 &  +\sqrt{3} \Big(32
   \sqrt{\Lt_k ^9 (3 \Lt_k +4)}-54 \sqrt{\Lt_k ^5 (3 \Lt_k +4)}+27 \sqrt{\Lt_k ^3 (3 \Lt_k +4)}\Big)+81 \Lt_k \Big) \Big) \, ,
\end{split}
\ee
\be
\begin{split}
p_2(\Lt_k) = & 5
   \Lt_k ^2 \Bigg(\eta  \Big(32 \Lt_k ^4-32 \Lt_k ^3-120 \Lt_k ^2+18 \Lt_k +27 \Big) \\
  &   +3 \Bigg(\frac{272 \Lt_k ^{7/2}}{\sqrt{17 \Lt_k +12}}-\frac{294
   \Lt_k ^{5/2}}{\sqrt{17 \Lt_k +12}}-\frac{513 \Lt_k ^{3/2}}{\sqrt{17 \Lt_k +12}}-16 \Lt_k ^3+90 \Lt_k ^2-81 \Lt_k \\
   & \qquad -54 \Big(3 \sqrt{\frac{\Lt_k }{17
   \Lt_k +12}}+1 \Big) \Bigg) \Bigg) \, ,
\end{split}
\ee
\be
\begin{split}
p_3(\Lt_k) = & 3 \Big(-13 \sqrt{\Lt_k ^{13} (13 \Lt_k +12)}-83 \sqrt{\Lt_k ^{11} (13 \Lt_k +12)}+51
   \sqrt{\Lt_k ^9 (13 \Lt_k +12)} \\
   & +494 \Lt_k ^7+99 \sqrt{\Lt_k ^7 (13 \Lt_k +12)}-272 \Lt_k ^6-672 \Lt_k ^5-54 \sqrt{\Lt_k ^5 (13 \Lt_k +12)} \\
   & \quad +468
   \Lt_k ^4+198 \Lt_k ^3-216 \Lt_k ^2 \Big) \, .
\end{split}
\ee

Note that despite their first-sight appearance, both  $\cS^{(0)}$ and $\cS^{(1)}$ are regular at $\Lt_k=1$, while they have a singularity at $\Lt_k=3/4$, which is a simple pole for $\cS^{(0)}$ and a pole of order two for $\cS^{(1)}$.
The presence of such poles is understood to be a generic feature of local truncations, disappearing in more generic truncations that better capture IR physics \cite{Machado:2007ea}.

\section{Fixed points and stability exponents}
\label{Sec:FPs}

From our on-shell equation in $d=4$ we find
\be \label{beta-tau}
\p_t \t_k  = \f{\t_k^2}{3\pi} \cS^{(0)} \equiv \b_\t(\t_k, \eta, \tG_k) \, ,
\ee
with  $\cS^{(0)}$ given by \eqref{S0}. As a consequence, the beta function for $\t_k$ is explicitly gauge-independent.
The explicit dependence on $\tG_k$ comes from the fact that $\cS^{(0)}$ is a function of $\Lt_k = \t_k/ \tG_k$ rather than of $\t_k$ alone.
Indeed we can actually write
\be
\b_\t(\t_k, \eta, \tG_k) = \tG_k^{2} f( \Lt_k,\eta) \, .
\ee
One then finds that there is a one-parameter family of zeros of $\b_\t$ with $\Lt^*=\Lt^*(\eta)$, but in order for any of these zeros to be a fixed point of $\t_k$, also $\tG_k$ must go to a fixed point.
And for $\tG_k$ to have a non-trivial fixed point, we must necessarily have $\eta=-2$. In this case the fixed point for the cosmological constant $\Lt_k$ is unique and gauge-independent!
This is the main result of our paper.
We have to use \eqref{eta-eq} in order to determine the value $\tG^*$ at which $\eta=-2$,  $\Lt^*=\Lt^*(\eta)$, and hence determine also the value of $\t^*$. The solution turns out to be unique, but gauge-dependent.
The same results can be obtained also by switching to the more usual picture (\ref{system2-a}-\ref{system2-b}). We find
\be \label{NGFP}
\Lt^* = 0.2612 \, , \,\,\,\,\, \tG^*=1.458 \, ,
\ee
corresponding to $\t^*=0.380968$. We stress again that this value of $\tG^*$ is obtained at $\a=\b=0$, while that of $\Lt^*$ is independent of such choice.
The computational difficulty is in keeping $\b$ generic, but for any fixed value of $\b$ the fixed point $\tG^*$ can be obtained as a function of $\a$ and $\eta_\a$. Varying $\b$ in the range $[-1,1]$, and with $\eta_\a=\eta$, the fixed-point value for Newton's constant was found to be typically of the form
$\tG^*(\a) = x/(y+w\, \a)$ with $x$, $y$ and $w$ real positive numbers of order one which depend on $\b$.
We see that $\tG^*(\a)$ goes to zero only for $\a\to\infty$ but such limit would mean no gauge-fixing, which is of course inconsistent; for $\a=-y/w$ we have a pole for $\tG^*$, and for smaller values Newton's constant becomes negative, but negative values of $\a$ should be excluded, as in the Euclidean path integral they amount to emphasizing the gauge degrees of freedom rather than suppressing them. Note also that  $\a= 0$ is expected to be a fixed point for the gauge parameter \cite{Ellwanger:1995qf,Litim:1998qi,Lauscher:2001ya}, hence negative values would never be reached for an initial condition at $\a>0$.

 An alternative fixed-point scenario, contemplated for example in \cite{Niedermaier:2006wt}, would be the one in which $\t_k$ reaches a fixed point while $\tG_k$ and $\Lt_k$ go one to infinity and the other to zero. We found that $\tG_k\to\infty$ is never a fixed point,\footnote{In the limit  $\tG_k\to\infty$ the function $\b_\t(\t_k, \eta, \tG_k)$ can go to zero only if $\eta=2$ and $\t_k\to0$. However, $\eta=2$ means $\tG_k\sim k^4$, hence $\t_k$ should be going to zero for $k\to\infty$; instead, solving \eqref{beta-tau} for $\eta=2$ we find that $\t_{k\to\infty}\to\infty$.} while $\tG_k\to 0$ is an IR fixed point with $\eta=0$ (and any finite value of $\t^*$, including zero, which corresponds to $\tG=\Lt=0$ and we will call the proper GFP).
However, the fixed points $(\t^*>0,\tG^*=0)$ are separated from the sector containing the proper GFP and the NGFP by the singularity at $\Lt_k=3/4$.
In conclusion, within the realm of validity of the Einstein-Hilbert truncation the only fixed points are the proper GFP and the NGFP \eqref{NGFP}.

 Around any fixed points  $\{\b_{g_i}(g_i^*)=0\}$ of the beta functions for the couplings $\{g_i\}$, the linearized RG flow $\p_t g_i = {\bf B}_{ij} (g_j - g^*_j)$ is governed by the stability matrix 
\be
{\bf B}_{ij} =  \p_j \beta_i {}_{|\{g_i^*\} }\ .
\ee
Defining the stability coefficients $\th_i$ as minus the eigenvalues of $\bf B$, the relevant (irrelevant) directions are associated to the eigenvectors
corresponding to stability coefficients with a positive (negative) real part.

The stability exponents are independent of the coupling parametrization, be it $(\t_k, \tG_k)$, $(\Lt_k,\tG_k)$ or any other (see \cite{Weinberg:1980gg} for a simple general proof), but they are mildly gauge-dependent. In the gauge  $\a=\b=0$ we find
\be
\th_\pm = 1.912 \pm 0.6685 \im \, .
\ee
The corresponding eigenvectors are of course parametrization-dependent, and in the  $(\Lt_k,\tG_k)$ plane we have
\be
V_\pm = \{ 0.2550 \pm 0.4958 \im, -0.8300 \} \, .
\ee

Comparing the fixed-point values and stability exponents found here with those found in the previous literature (as collected for example in Table~1 of \cite{BMS2}), we can observe 
a qualitative (and grosso modo quantitative) agreement with previous results. In particular, the $\a$-dependence of $\tG^*$ is very similar to that found in \cite{Souma:2000vs}.
The main novelty of our result is that all the gauge-dependence is carried by Newton's constant, while the fixed-point value of the cosmological constant is gauge-independent.

\section{Discussion and outlook}
\label{Sec:concl}

We have achieved the slightly paradoxical conclusion that whereas $\t_k$ is the only essential parameter in the Einstein-Hilbert truncation, its value at the non-Gaussian fixed point is as gauge-dependent as those of $\tG_k$, 
while the NGFP value of $\Lt_k$ is gauge-independent, and given in $d=4$ by $\Lt^*=0.2612$.
The reason behind such outcome is that, while the beta function of $\t_k$ has no explicit gauge-dependence, it is a function of $\tG_k$, whose flow is instead governed by a gauge-dependent function. Nevertheless the particular dependence of $\b_\t$ on the combination $\t_k/\tG_k=\Lt_k$ implies a gauge-independent fixed point value for $\Lt_k$,
in agreement with one-loop results \cite{Niedermaier:2010zz}.

The appearance of $\tG_k$ in the beta function of $\t_k$ is a practical consequence of the special role of the metric, for which the scaling of the field and the scaling of the momenta are the same operation, as discussed in \cite{Percacci:2004sb}. 
Another way to understand the peculiar role of Newton's constant might be to note that
$G$ can only be removed from the action by a redefinition of the full metric $g_{\m\n}+h_{\m\n}$, or by a redefinition of the graviton field $h_{\m\n}$ with a field-independent shift.
Both redefinitions are unusual and it is not obvious that they are contemplated by standard proofs of equivalence of S-matrix elements.

It should be noted that the same phenomenon can be observed in one-loop calculations for the running of $\t_k$ \cite{Niedermaier:2010zz}, as well as of other essential parameters \cite{Benedetti:2011nd}. In such cases one can observe that it is the contribution of quadratic divergencies that is bringing Newton's constant into the beta functions of the essential parameters, highlighting the fact that the dimensional nature of $G$ is another peculiar aspect of the game.

As the location of a fixed point is not expected to be a universal property, its gauge-dependence might not be an issue, at least as long the fixed point is found in any viable gauge. On the other hand, critical exponents are usually universal, and their gauge-dependence appears as an annoyance. 
In this respect, it would be interesting to study the running of the gauge parameters as well: we already have arguments \cite{Ellwanger:1995qf,Litim:1998qi} supporting a fixed point at $\a=0$; it would be interesting to confirm that for gravity and look also for fixed points of $\b$. Despite the unphysical role of the gauge-fixing we cannot exclude that the gauge parameters would be forced to go to their fixed-point value at the gravitational fixed point.

It is interesting to point out that in the scheme we presented here, and because of the restriction to a maximally symmetric background, the truncation method has a clear physical interpretation:
higher order terms are further off-shell and might hence be discarded. This might explain the nice convergence properties of the polynomial truncations in $R$ \cite{Codello:2008vh,Machado:2007ea}.
On the other hand, on a non-spherical background one would have no reason to discard $(C_{\m\n\r\s})^3$ and higher powers of the Weyl tensor, as in general this is not zero on shell.
However, as soon as we add any other of such terms to the truncation the equations of motion change
(remember that we are not using perturbation theory, and the equations of motion are those of the effective action, not of the bare one).

Finally we should mention that even though we concentrated here on the Einstein-Hilbert truncation for gravity, the ideas and methods introduced could be applied also to other truncations as well as to other gauge theories. A more general treatment and further applications will be the subject of future work.

\subsection*{Acknowledgements}
%
I would like to thank Roberto Percacci, Oliver Rosten and Simone Speziale for useful conversations on some of the topics discussed here.


\providecommand{\href}[2]{#2}\begingroup\raggedright\endgroup

\end{document}